\documentclass[10pt]{article}

\usepackage{amsmath,amsfonts,amssymb,amsthm,graphicx}
\usepackage{enumerate}
\usepackage{color}
\usepackage[colorlinks=true,citecolor=blue,pdfpagemode=UseNone,pdfstartview=FitH]{hyperref}

\allowdisplaybreaks[4]

\DeclareMathOperator{\VS}{VS}

\theoremstyle{plain}
\newtheorem{theorem}{Theorem}

\theoremstyle{definition}

\theoremstyle{remark}
\newtheorem{remark}[theorem]{Remark}

\title{A note on data splitting with e-values:
  online appendix to my comment on Glenn Shafer's ``Testing by betting'' \cite{Shafer:2020-local}}
\author{Vladimir Vovk\thanks%
  {Department of Computer Science,
  Royal Holloway, University of London,
  Egham, Surrey, UK.
  E-mail: \href{mailto:v.vovk@rhul.ac.uk}{v.vovk@rhul.ac.uk}.}}

\begin{document}
\maketitle

\begin{abstract}
  This note reanalyzes Cox's idealized example of testing with data splitting
  using e-values (Shafer's \cite{Shafer:2020-local} betting scores).
  Cox's exciting finding was that the method of data splitting,
  while allowing flexible data analysis,
  achieves quite high efficiencies, of about 80\%.
  The most serious objection to the method was that it involves splitting data at random,
  and so different people analyzing the same data may get very different answers.
  Using e-values instead of p-values remedies this disadvantage.

  The version of this note at \url{http://alrw.net/e} (Working Paper 7)
  is updated most often.
\end{abstract}

\section{Introduction}
\label{sec:introduction}

Data splitting is a simple method for performing hypothesis testing
using procedures chosen in the light of the data.
The data is randomly split into disjoint parts,
some of which used for choosing a test and another for performing it.

The method was analysed computationally and theoretically by Cox \cite{Cox:1975} in 1975
and further discussed in his 1977 review \cite[Section 3.2]{Cox:1977},
where he describes the method as well known and refers to an \emph{American Statistician} paper \cite{Selvin/Stuart:1966}
with a wide-ranging discussion of ``snooping'', ``fishing'', and ``hunting'' in data analysis.

As summarized in \cite[Section 3.2]{Cox:1977},
Cox \cite{Cox:1975} had analyzed the method of data splitting theoretically in a simple idealized situation
showing that it achieves quite high efficiencies, of about 80\%.
However, Cox \cite{Cox:1975} also points out an obvious disadvantage of the method:
\begin{quote}
  Any method involving randomization in analysis
  is such that different investigators analyzing the same data by nominally the same method get different answers.
\end{quote}
This criticism is reiterated in \cite{Cox:1977}, including the discussion.

Using e-values \cite{Vovk/Wang:arXiv1912a-local} (also know as betting scores \cite{Shafer:2020-local})
makes the problem of getting different answers by different investigators much less serious.
The average of e-values is always an e-value,
and so we can average e-values resulting from several random data splits
obtaining a less random valid e-value
(an almost deterministic one if the number of splits is very large).
This note studies the dependence of the resulting p-values in Cox's idealized situation on the random split of the data
and then performs similar (but simpler) analysis using e-values instead of p-values.

For a recent review of testing by data splitting, see \cite[Section 1]{DiCiccio/etal:2020}.
That paper also discusses methods based on combining p-values resulting from different data splits,
both using the median (Section 2.1) and the arithmetic mean (Section 2.2).
Both methods of combination, however, result in an extra factor of 2
(which has served as an inspiration for this note).
Combination of e-values by averaging is discussed, in a similar context,
in \cite[Section 4]{Wasserman/etal:2020},
which refers to averaging a large number of e-values as derandomization.

Section~\ref{sec:p} discusses the dependence on the random data split of the resulting p-values in Cox's \cite{Cox:1975} setting
(discussed in Section~\ref{sec:Cox}).
We will see that in many cases it is significant.
Section~\ref{sec:e} replaces p-values by e-values.
To compare e-values and p-values,
we use Shafer's \cite{Shafer:2020-local} calibrator
and Jeffreys's rule of thumb \cite[Appendix~B, p.~435]{Jeffreys:1961}:
a p-value of $0.05$ corresponds to an e-value of $10^{1/2}$,
and a p-value of $0.01$ corresponds to an e-value of $10$.

This note is not self-contained in that it does not define e-values;
see \cite{Shafer:2020-local} or \cite{Vovk/Wang:arXiv1912a-local} for definitions.

\section{Cox's ideal situation}
\label{sec:Cox}

We are given $m$ independent random samples of size $r$
from normal populations with means $\mu_1,\dots,\mu_m$ and known common variance $\sigma_0^2$.
The null hypothesis is that all means are zero,
and the alternative is that just one of the means is positive, $\mu>0$.
We apply the method of data splitting by dividing each sample into two portions of sizes $p r$ and $(1-p)r$.
We then take the population for which the first-portion sample mean is largest.
Finally we apply the standard one-sided normal test to the mean of the corresponding second portion,
ignoring the second-portion samples of the other $m-1$ populations.

Cox defines the \emph{effective} level $\alpha$ power of this data-splitting procedure
to be the probability that the correct population is chosen
and that the second-portion sample is significant at least at level $\alpha$.
It is given by the formula
\begin{equation}\label{eq:Cox-split-power}
  \Phi
  \left(
    -k^*_{\alpha} + \Delta\sqrt{1-p}
  \right)
  \int_{-\infty}^{\infty}
  \Phi(v)^{m-1}
  \phi
  \left(
    v-\Delta\sqrt{p}
  \right)
  dv
\end{equation}
where $\Delta := \mu\sqrt{r}/\sigma_0$,
and $\phi$ and $\Phi$ are, respectively, the standard normal density function and distribution function,
and $k^*_{\alpha}$ is defined by $\Phi(-k^*_{\alpha}) = \alpha$.

Cox also defines an \emph{exact procedure} that tests the means collectively for significance
using the largest mean as test statistic.
(Cox often surrounds ``exact'' by quotes, but let us omit those.)
If the largest mean is significant at level $\alpha_m$ in the usual test for a single mean,
its level of significance after allowing for selection is
\begin{equation}\label{eq:alpha-m}
  \alpha
  :=
  1 - (1 - \alpha_m)^m.
\end{equation}
The effective level $\alpha$ power (where ``effective'' is used in the same sense of choosing the correct population) is
\begin{equation}\label{eq:Cox-exact-power}
  \int_{k^*_{\alpha_m}}^{\infty}
  \Phi(v)^{m-1}
  \phi
  \left(
    v-\Delta
  \right)
  dv,
\end{equation}
where $\alpha_m$ is defined by \eqref{eq:alpha-m}.

\begin{table}
  \!\!\!\!\!\!
  \begin{tabular}{cc|cccc|cccc}
    && \multicolumn{4}{c|}{$\alpha=0.1$} & \multicolumn{4}{c}{$\alpha=0.01$} \\
    $m$ & $\Delta$ & $p=0.2$ & $p=0.4$ & $p=0.6$ & exact & $p=0.2$ & $p=0.4$ & $p=0.6$ & exact \\
    \hline
      2 &        1 &    0.22 &    0.21 &    0.18 &  0.26 &   0.047 &   0.041 &   0.032 & 0.058 \\
        &        2 &    0.31 &    0.49 &    0.43 &  0.64 &    0.22 &    0.18 &    0.12 &  0.28 \\
	&        4 &    0.89 &    0.93 &    0.88 &  0.99 &    0.80 &    0.75 &    0.57 &  0.92 \\
    \hline
     10 &        1 &   0.065 &   0.071 &   0.070 & 0.092 &   0.014 &   0.014 &   0.012 & 0.018 \\
        &        2 &    0.21 &    0.26 &    0.26 &  0.37 &   0.091 &   0.094 &   0.076 &  0.14 \\
	&        4 &    0.60 &    0.79 &    0.82 &  0.95 &    0.54 &    0.64 &    0.53 &  0.82 \\
	&        6 &    0.85 &    0.97 &    0.99 & 1.000 &    0.85 &    0.96 &    0.93 & 0.998
  \end{tabular}
  \caption{Cox's results for the effective power of data splitting for different $p$
    and the effective power of the exact test}
  \label{tab:Cox-results}
\end{table}

Notice that \eqref{eq:Cox-split-power} and \eqref{eq:Cox-exact-power} are only lower bounds for the power as it is usually defined.
Cox's numeric results about the effective power of these tests are given in Table~\ref{tab:Cox-results}.
In our experiments we will always use $p:=0.4$ (the middle value).

\section{Experiments 1: variability of p-values}
\label{sec:p}

Let us assume, without loss of generality, that $\sigma_0=1$.
The results of our computational experiments are very much affected by the choice of the random seed
for the random number generator (but our conclusions will not be affected, of course).

The \emph{data-split p-value} is computed as
\begin{equation}\label{eq:p-data-split}
  k^*_{\alpha},
\end{equation}
where $\alpha:=b/\sigma=\sqrt{(1-p)r}b$,
$b$ is the mean of the second portion of the sample with the largest mean of the first portion,
and
\begin{equation}\label{eq:sigma}
  \sigma
  :=
  1/\sqrt{(1-p)r}
\end{equation}
is the standard deviation of the second portion of each sample.

Similarly, the \emph{exact p-value} is computed as
\begin{equation}\label{eq:p-exact}
  1-(1-k^*_{\alpha})^m,
\end{equation}
where $\alpha:=\sqrt{r}c$ and $c$ is the largest mean of all $m$ samples.

\begin{figure}
  \begin{center}
    \includegraphics[width=0.6\textwidth]{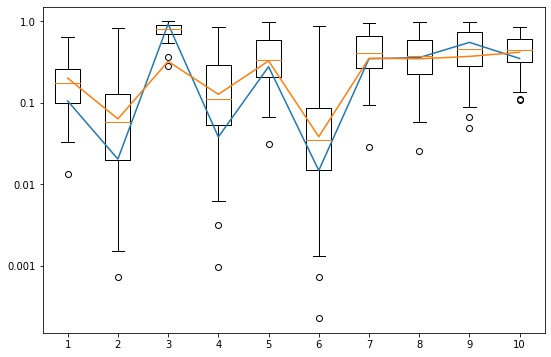}
  \end{center}
  \caption{Box plots for 10 randomly generated datasets with $m=2$ and $\Delta=1$,
    with the exact p-values added in blue (details in text)}
  \label{fig:m2_Delta1}
\end{figure}

Results for one of the low-power cases (according to Table~\ref{tab:Cox-results})
are shown as Figure~\ref{fig:m2_Delta1}.
The 10 boxplots correspond to 10 datasets (each consisting of $m=2$ samples)
randomly generated from the normal distribution $N(\mu,1)$ making $\Delta=1$.
The value of $r$ (sample size) is set to 100 (see Remark~\ref{rem:r} below);
therefore, $\mu=\Delta/\sqrt{r}=0.1$.
The exact p-values for the same 10 datasets are shown in blue, and for visibility they are connected with blue lines.
(The orange lines should be ignored in this section; they will be explained in Section~\ref{sec:e}.)
Each dataset is split randomly 100 times, and the corresponding 100 p-values \eqref{eq:p-data-split}
are summarized as boxplot (whose box is bounded by the quartiles and contains the median;
for the rules governing the whiskers see the \texttt{matplotlib} documentation).

\begin{remark}\label{rem:r}
  The role of $r$ is not essential in Cox's calculations as long as $p r$
  (the size of the first portion of each sample)
  is integer (in the context of Table~\ref{tab:Cox-results}, as long as $r$ is divisible by 5).
  In particular, $r$ does not enter \eqref{eq:Cox-split-power} or \eqref{eq:Cox-exact-power}.
  In our experiments, however, we can't set $r:=5$ since this would lead
  to only $\binom{5}{2}=10$ possible data splits for each sample,
  which would show in the boxplots, especially for $m=2$.
\end{remark}

\begin{figure}
  \begin{center}
    \includegraphics[width=0.6\textwidth]{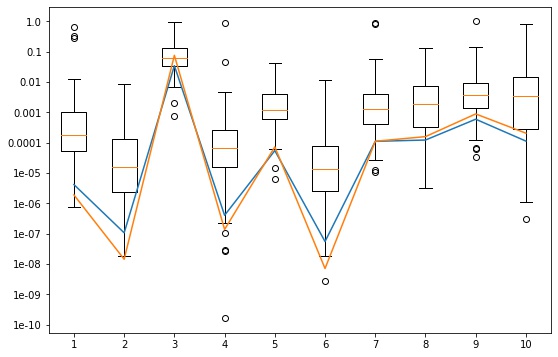}
  \end{center}
  \caption{The analogue of Figure~\ref{fig:m2_Delta1} for $m=2$ and $\Delta=4$}
  \label{fig:m2_Delta4}
\end{figure}

\begin{figure}
  \begin{center}
    \includegraphics[width=0.6\textwidth]{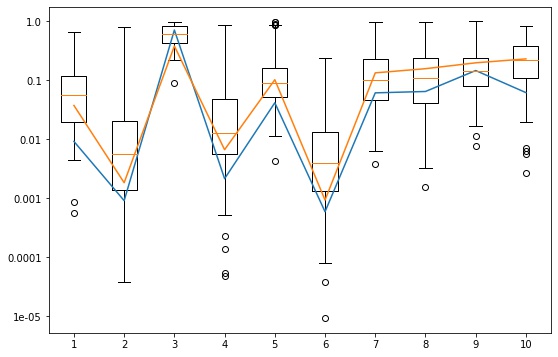}
  \end{center}
  \caption{The analogue of Figure~\ref{fig:m2_Delta1} for $m=2$ and $\Delta=2$}
  \label{fig:m2_Delta2}
\end{figure}

A high-power case is shown as Figure~\ref{fig:m2_Delta4}.
Most of the results are highly statistically significant (the p-values are below $1\%$),
and so, from the point of view of the conventional levels $1\%$ and $5\%$,
it is not as informative as the mid-power case shown as Figure~\ref{fig:m2_Delta2}.
In that figure, the statistical significance of data-split p-values
strongly depends on the random data split.

\begin{figure}
  \begin{center}
    \includegraphics[width=0.48\textwidth]{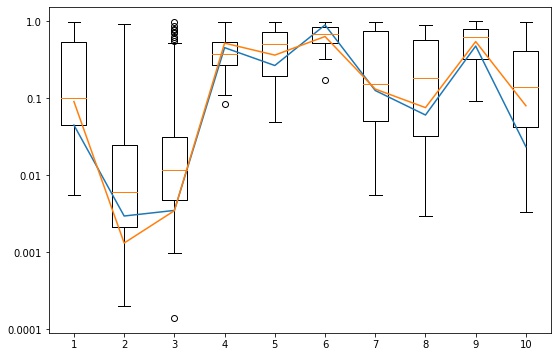}
    \includegraphics[width=0.48\textwidth]{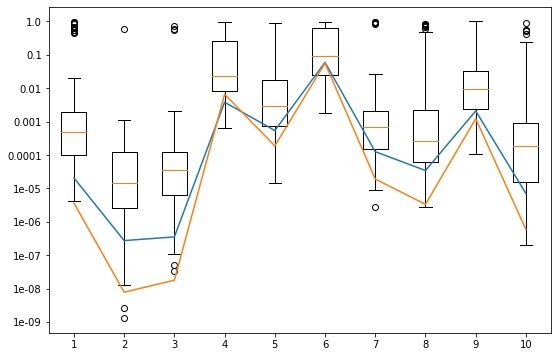}
  \end{center}
  \caption{The analogues of Figure~\ref{fig:m2_Delta1} for $m=10$ (both panels)
    and $\Delta=2$ (left panel) and $\Delta=4$ (right panel)}
  \label{fig:m10-middle}
\end{figure}

Figure~\ref{fig:m10-middle} reports results for $m=10$ and $\Delta\in\{2,4\}$
(the two mid-power cases).
The effect of the randomness in the data split on the statistical significance of the resulting p-values
is again substantial.

\section{Experiments 2: using e-values}
\label{sec:e}

There is a natural way to define data-split e-values.
As Shafer \cite{Shafer:2020-local} discusses in Section 2.2,
a natural choice is the likelihood ratio between an alternative and the null hypotheses.
The true alternative (namely, its parameter $\mu$) is not known,
but in Cox's data-splitting scheme we can use the largest mean $a$ of the first portion as an estimate of $\mu$,
which gives us the likelihood ratio
\begin{equation}\label{eq:e}
  e
  :=
  \frac
  {(2\pi)^{-1/2}\exp\left(-\frac{(b-a)^2}{2\sigma^2}\right)}
  {(2\pi)^{-1/2}\exp\left(-\frac{b^2}{2\sigma^2}\right)}
  =
  \exp\left(\frac{ab}{\sigma^2} - \frac{a^2}{2\sigma^2}\right),
\end{equation}
where $b$ is the mean of the second portion of the sample with the largest mean $a$ of the first portion,
and $\sigma$ is the standard deviation \eqref{eq:sigma} of the mean of the second portion.

It is not immediately clear, however, how to define an e-value analogue for Cox's exact p-values.
Not knowing the alternative hypothesis hurts more in the case of e-values;
while the Neyman--Pearson p-values do not depend on the parameter $\mu$,
we cannot set the e-value to the likelihood ratio since it does depend on $\mu$.
A natural way out is to choose a ``prior distribution'' over $\mu$
and mix the likelihood ratios corresponding to different $\mu$ with respect to that distribution,
but this appears more ad hoc and more complicated than what we did in the case of data splitting.

\begin{figure}
  \begin{center}
    \includegraphics[width=0.6\textwidth]{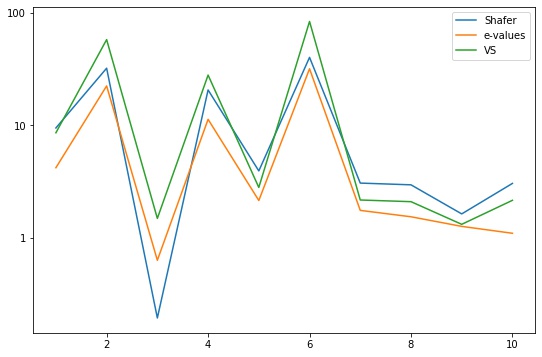}
  \end{center}
  \caption{The average e-values for $m=2$ and $\Delta=2$ in orange
    complemented by calibrated p-values in blue and green}
  \label{fig:e_m2_Delta2}
\end{figure}

Figure~\ref{fig:e_m2_Delta2} presents the e-values \eqref{eq:e} for $m=2$ and $\Delta=2$;
therefore, it contains some common information with Figure~\ref{fig:m2_Delta2},
which is, however, presented differently.
We adapt the method of \cite[Section 1]{DiCiccio/etal:2020} to e-values
(as already mentioned in Section~\ref{sec:introduction}).
Instead of the boxplots of p-values in Figure~\ref{fig:m2_Delta2},
we can now average the 100 e-values \eqref{eq:e} for $m=2$ and $\Delta=2$,
obtained from the same datasets and the same data splits as in Figure~\ref{fig:m2_Delta2}
(the remaining randomness in the average e-values is tiny
and boxplots would not be informative for them).
The average e-values are shown in orange in Figure~\ref{fig:e_m2_Delta2}.
Using Jeffreys's \cite[Appendix~B]{Jeffreys:1961} rule of thumb already mentioned in Section~\ref{sec:introduction},
we can regard e-values above $10^{1/2}$ as statistically significant
and e-values above $10$ as highly statistically significant.
We can see that two or three resulting e-values are highly statistically significant in this sense.

\begin{remark}
  We could eliminate \emph{all} randomness in the average e-values
  by averaging over all splits of the samples into first and second portions,
  as discussed in \cite[Section 4]{Wasserman/etal:2020}.
  This would be feasible for $m=2$ and $r=5$ (with 10 possible splits for each sample)
  but not in our case $r=100$ (with about $10^{65}$ possible splits for each sample).
\end{remark}

As Shafer says in section ``Comparing scales'' \cite[Section 3]{Shafer:2020-local},
there is no one way to compare p-values, such as \eqref{eq:p-exact},
and e-values, such as \eqref{eq:e}.
As discussed in detail in \cite{Vovk/Wang:arXiv1912a-local},
there are various ways of \emph{calibrating} p-values, i.e., making them into e-values
(although there is essentially one way, $e\mapsto1/e$, of making e-values into p-values).

\begin{table}
  \begin{center}
  \begin{tabular}{ccccc}
    p-value $p$ & $1/p$ & $S(p)$ & Jeffreys & $\VS(p)$ \\
    \hline
    1 & 1 & 0 & & 1\\
    0.1 & 10 & 2.2 & & 1.6 \\
    0.05 & 20 & 3.5 & 3.2 & 2.5 \\
    0.01 & 100 & 9.0 & 10 & 8.0 \\
    0.005 & 200 & 13.1 & & 13.9 \\
    0.001 & 1,000 & 30.6 & & 53.3 \\
    0.000001 & 1,000,000 & 999 & & 26628
  \end{tabular}
  \end{center}
  \caption{Calibrating p-values}\label{tab:calibrating}
\end{table}

Let us first apply Shafer's \emph{calibrator}
(i.e., function transforming p-values into e-values)
\begin{equation}\label{eq:Shafer}
  S(p)
  :=
  \frac{1}{\sqrt{p}} - 1
\end{equation}
\cite[(6)]{Shafer:2020-local}
to Cox's exact p-values.
Table~\ref{tab:calibrating}
(an extension of Shafer's \cite{Shafer:2020-local} Table 2 ``Making a p-value into a betting score'')
gives the e-values produced by Shafer's calibrator in the third column
and Jeffreys's \cite[Appendix B]{Jeffreys:1961} estimates,
which are remarkably close to Shafer's.

The exact p-values shown in blue in Figure~\ref{fig:m2_Delta2}
are shown in Figure~\ref{fig:e_m2_Delta2}, also in blue,
after being transformed into e-values by Shafer's calibrator \eqref{eq:Shafer}.
They are fairly close to the orange lines.
Notice that in one respect the comparison between the blue and orange lines is unfair to the average e-values:
the input p-values are exact while the average e-values are based on data splitting.
(On the other hand, our treatment of p-values and e-values is not symmetric
in that we compare them in the e-domain,
which makes sense in discussing a paper promoting betting scores
as language for statistical and scientific communication.)

The orange lines in the previous figures, Figures~\ref{fig:m2_Delta1}--\ref{fig:m10-middle},
show the average e-values in the p-domain using the inverse transformation $S^{-1}$ to Shafer's calibrator
\[
  S^{-1}(e)
  =
  (e+1)^{-2}.
\]
Of course, these are not valid p-values;
they are the p-values we would need to obtain our average e-values
by using Shafer's calibrator.
We can see that for all those figures we obtain similar e-values
directly by using data splitting and by applying Shafer's calibrator to exact p-values.

A particularly natural class of calibrators is
\begin{equation}\label{eq:Sellke}
  p
  \mapsto
  \epsilon p^{\epsilon-1},
\end{equation}
where $\epsilon\in(0,1)$ is the parameter.
For a small $\epsilon$, it comes close to the ideal (but not attainable) calibrator $p\mapsto1/p$
(if we ignore constant factors, as customary in the algorithmic theory of randomness,
an area where many of these ideas originated).
To get rid of the parameter $\epsilon$, we may consider the upper bound
\[
  \VS(p)
  :=
  \sup_{\epsilon}
  \epsilon p^{\epsilon-1}
  =
  \begin{cases}
    -\exp(-1)/(p\ln p) & \text{if $p\le\exp(-1)$}\\
    1 & \text{otherwise}.
  \end{cases}
\]
It was proposed in \cite[Section 9]{Vovk:1993logic} and, independently, \cite{Sellke/etal:2001}.
It should be remembered that it is not a valid e-value
and just shows what is attainable with the calibrators in the class \eqref{eq:Sellke}.

Figure~\ref{fig:e_m2_Delta2} also shows the VS transformations of the exact p-values (in green).
As expected, they are slightly above the bona fide e-values shown in orange and blue.
The VS values are also given in Table~\ref{tab:calibrating} (last column).
We can see that the VS transformation is also not so different from Shafer's calibrator and Jeffreys's intuition
unless the input p-value is very large or extremely small.

\begin{figure}
  \begin{center}
    \includegraphics[width=0.6\textwidth]{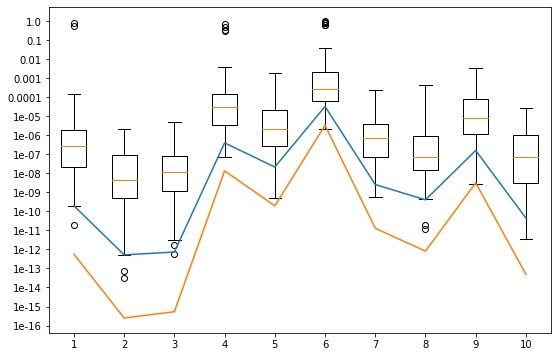}
  \end{center}
  \caption{The analogue of Figure~\ref{fig:m2_Delta1} for $m=10$ and $\Delta=6$}
  \label{fig:m10_Delta6}
\end{figure}

We can see already in Figure~\ref{fig:m10-middle} that for a strong signal
the performance of e-values tends to improve as compared with Shafer-calibrated p-values.
This is illustrated further by Figure~\ref{fig:m10_Delta6}, where $\Delta=6$.
The e-values produced by Shafer's calibrator are significantly worse
than the e-values obtained by data splitting
(which has a whiff of superefficiency).

\begin{figure}
  \begin{center}
    \includegraphics[width=0.6\textwidth]{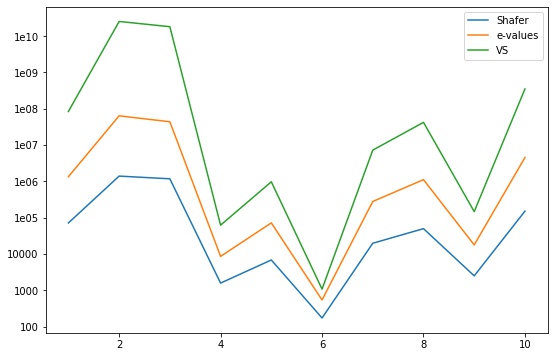}
  \end{center}
  \caption{Blue and orange lines of Figure~\ref{fig:m10_Delta6} in the e-domain
    complemented by the VS plot}
  \label{fig:e_m10_Delta6}
\end{figure}

Figure~\ref{fig:e_m10_Delta6} shows the blue and orange lines of Figure~\ref{fig:m10_Delta6} in the e-domain,
complemented by the VS bound.
We can see that using some calibrators in the class \eqref{eq:Sellke}
allows us to beat the e-values obtained by data splitting.
(The problem, however, is that we do not know how to choose the right $\epsilon$ in advance.)

\section{Conclusion}

Despite the high efficiency of the method of data splitting applied to p-values
(at least in Cox's idealized situation),
the adoption of the method may have been hindered by the significant amount of randomness at the stage of data analysis,
with different analysts potentially arriving at very different conclusions while using the same approach.
A byproduct of adopting the language of betting and e-values as a means of statistical and scientific communication
\cite{Shafer:2020-local}
is that this drawback disappears.

\subsection*{Acknowledgments}

Many thanks to Ruodu Wang for his advice.
The computational experiments in Sections~\ref{sec:p}--\ref{sec:e} have been performed in Python;
the Jupyter notebook containing the code is available on arXiv (go to ``Other formats'').
This research has been partially supported by Amazon, Astra Zeneca, and Stena Line.

\appendix
\section{The distribution of p-values under the alternative hypothesis}

It is customary in statistical hypothesis testing
to talk about the size and power of statistical tests
in the toy situation of a simple null hypothesis and a simple alternative
(see, e.g., \cite[Section 3.1]{Lehmann/Romano:2005}).
This was essentially the language of Cox's paper \cite{Cox:1975} and Section~\ref{sec:Cox},
except that power was replaced by its lower bound, effective power.
The experiments of Section~\ref{sec:p}, however,
concerned the behaviour of the p-values under the alternative hypothesis.
This appendix will spell out the connections and explore the distribution of p-values experimentally.

The distribution function $F$ of the p-values under the alternative hypothesis
is exactly the power as function of the test size.
Let us check this, assuming that the p-value is distributed uniformly on $[0,1]$
under the null hypothesis.
For any significance level $\alpha\in(0,1)$,
the critical region of size $\alpha$ consists of all observations producing a p-value $p\le\alpha$.
Its probability $F(\alpha)$ is, by definition, the power at significance level $\alpha$.

\begin{figure}
  \begin{center}
    \includegraphics[width=0.6\textwidth]{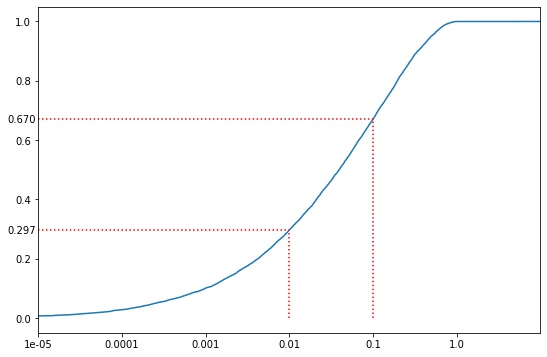}
  \end{center}
  \caption{The distribution function of exact p-values for $m=2$ and $\Delta=2$}
  \label{fig:p}
\end{figure}

Figure~\ref{fig:p} gives the distribution function of exact p-values for $m=2$ and $\Delta=2$,
which are the values in Table~\ref{tab:Cox-results} that have the effective powers
farthest from the end-points of the interval $[0,1]$.
The powers corresponding to the sizes $\alpha\in\{0.1,0.01\}$
are greater than the effective powers reported in Table~\ref{tab:Cox-results}
($0.670>0.64$ and $0.297>0.28$), but the difference is modest.
To plot the distribution function in Figure~\ref{fig:p},
I have generated from the alternative distribution for the data
10,000 exact p-values (valid under the null distribution).
\end{document}